\documentclass[twocolumn,10pt,pra,showpacs]{revtex4}
\usepackage{amsmath,graphicx,amsfonts,bm,amssymb,times}
\begin{document}
\title{Simple unconventional geometric scenario  of one-way quantum
computation with superconducting qubits inside a cavity}
\author{Zheng-Yuan Xue}
\author{Z. D. Wang}
\email{zwang@hkucc.hku.hk}

\affiliation{Department of Physics and Center of Theoretical and
Computational Physics, The University of Hong Kong, Pokfulam Road,
Hong Kong, China}

\date{\today}

\begin{abstract}
We propose a simple unconventional geometric scenario to achieve a
kind of nontrivial multi-qubit operations with superconducting
charge qubits placed in a microwave cavity. The proposed quantum
operations are insensitive not only to the thermal state of cavity
mode but also to certain random operation errors, and thus may lead
to high-fidelity quantum information processing. Executing the
designated quantum operations, a class of highly entangled cluster
states may be generated efficiently in the present scalable
solid-state system, enabling one to achieve one-way quantum
computation.
\end{abstract}


\maketitle

Quantum computers have been paid much attention for the past decade,
as they may accomplish certain tough tasks that can hardly be
fulfilled by their classical counterpart. Despite rather advanced
theoretical concepts of quantum computation, practical physical
implementation appears to be at an early stage. Recently,
superconducting qubits have attracted significant interests because
of their potential suitability for scalable quantum computation
\cite{Makhlin}. The realization of an entangled state of two qubits
\cite{fluxent,chargeent} and implementation of a conditional phase
gate operation \cite{cg} were reported in this scalable solid state
system. Note that, superconducting qubits are quite sensitive to the
external environment and the background noise, with the decoherence
time being rather short. In order to couple multipartite qubits,  a
lot of auxiliary devices are often needed, which increases the
complexity of circuits and also inevitably introduces additional
uncontrollable noises, and thus would make the fidelity and
scalability of the system no longer better. On the other hand, the
cavity quantum electrodynamics (QED), which addresses the properties
of atoms coupled to discrete photon modes in high $Q$ cavities, was
also proposed as a potential setup for quantum information
processing including quantum computation \cite{qedreview}. In sharp
contrast to the above resource-consuming coupling, the cavity mode
can act as a "bus" and thus easily mediate a kind of long range
interactions among the superconducting qubits
\cite{Almaas,kis,Wallraff,liu,zhucavity}. In this sense, an idea to
place superconducting qubits in the cavity (i. e., the
superconducting cavity QED) is more promising for quantum
computation, being a macroscopic analogy of atomic quantum computing
and control. This quantum setup not only provides strong inhibition
of spontaneous emission, which leads to the enhancement of qubit
lifetimes, but also suppresses greatly the decoherence caused by the
external environment since the cavity may serve as a magnetic
shield.

In this paper, we propose a simpler scheme  for implementing a kind
of unconventional geometric phase gates \cite{zhuunconventional}
with superconducting charge qubits coupled to a microwave cavity
mode \cite{zhucavity}. The proposed quantum operations depend only
on global geometric features \cite{zhuwang} and are insensitive to
the thermal state of cavity modes. Thus they may lead to the
high-fidelity quantum information processing. More importantly,
executing the designated quantum operations, a class of highly
entangled cluster states may be generated efficiently and thus a new
kind of one-way quantum computation scheme can be achieved.

\begin{figure}[bp]
\includegraphics[scale=0.5,angle=0]{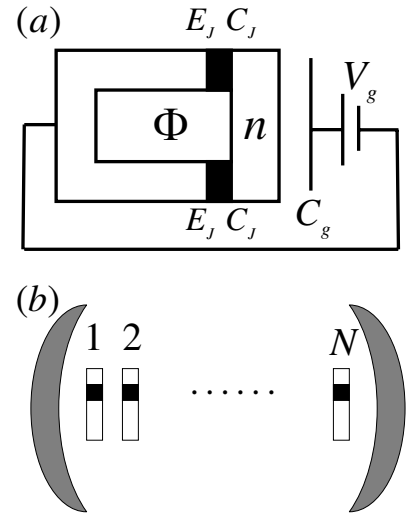}
\caption{(a) A schematic superconducting charge qubit connected to a
SQUID loop consisting of two identical Josephson junctions, subject
to a controllable gate voltage $V_g$  and a magnetic flux $\Phi$.
(b) $N$ qubits are placed in a microwave cavity, where the
qubit-qubut couplings are mediated by the cavity mode.} \label{fig1}
\end{figure}

A single superconducting qubit considered here, as shown in Fig
(\ref{fig1}a), consists of a small superconducting box with excess
Cooper-pair charges, formed by an symmetric superconducting quantum
interference device (SQUID) with the capacitance $C_{J}$ and
Josephson coupling energy $E_J$, pierced by an external magnetic
flux $\Phi$. A control gate voltage $V_g$ is connected to the system
via a gate capacitor $C_g$. The Hamiltonian of the system is
\cite{Makhlin}
\begin{equation}
\label{h1} H=E_{c}(n-\bar{n})^2-E_J\cos\varphi_1-E_J\cos\varphi_2,
\end{equation}
where $n$ is the number operator of (excess) Cooper-pair charges
on the box, $E_{c}=2e^2/(C_g+2C_{J})$ is the charging energy,
$\bar{n}=C_g V_g/2$ is the induced charge  controlled by the gate
voltage $V_g$, and $\varphi_m$ ($m=1, 2$) is the gauge-invariant
phase difference between the two sides of the $m$th junction. We
here focus on the charging regime where $E_{c}\gg E_{J}$. In this
case,  a convenient basis is formed by the charge states,
parameterized by the number of Cooper pairs $n$ on the box with
its conjugate $\varphi$; they satisfy the standard commutation
relation: $[\varphi, n]=i$. At temperatures much lower than the
charging energy and restricting the gate charge to the range of
$N_g \in [0,1]$, only a pair of adjacent charge states
$\{|0\rangle,|1\rangle\}$ on the island are relevant. The
Hamiltonian (\ref{h1}) is then reduced to
\begin{equation}
\label{h1r} H=-E_{ce}\sigma_z-E(\Phi)\sigma_x,
\end{equation}
where $E_{ce} =2E_c(1-2\bar{n})$,
$E(\Phi)=E_J\cos(\pi\frac{\Phi}{\phi_0})$, $\sigma_x$ and $\sigma_z$
are the Pauli matrices. Clearly, two noncommuting single-qubit gates
$\sigma_x$ and $\sigma_z$ can be obtained directly from this
Hamiltonian by simply tuning the gate voltage close to the
degeneracy point ($\bar{n}\sim 1/2$) or adjusting the external flux
$\Phi$. For a system of $N$ independent qubits with each being at
the degeneracy point and $\Phi$ being time-independent, the time
evolution of such system may be written as
\begin{equation}
\label{u1}
U(t)_1=\exp\left[i\frac{E(\Phi)}{\hbar}t\sum_{j=1}^N\sigma_x^j\right].
\end{equation}

On the other hand, let us consider a superconducting qubit to be
placed in a cavity as shown in Fig. (\ref{fig1}b). The
gauge-invariant phase difference is
$$\varphi_m^{'}=\varphi_m-\frac{2\pi}{\phi_0}\int_{l_m} {\bf A}_m
\cdot d {\bf l}_m,$$ where  ${\bf A}_m$ is the vector potential in
the same gauge of $\varphi_m$. ${\bf A}_m$ may be divided into two
parts ${\bf A}_m^\prime+{\bf A}_m^\phi $, where the first and
second terms arise respectively from the electromagnetic field of
the cavity normal modes and the external magnetic flux,
respectively. For simplicity, we here assume that the cavity has
only a single mode to play a role. In the Coulomb gauge, ${\bf
A}_m^\prime $ takes the form $\sqrt{\hbar/2 \omega_c
V}(a+a^{\dagger})\hat{\epsilon}$ \cite{Almaas}, where
$\hat\epsilon$ is the unit polarization vector of  cavity mode,
$V$ is the volume of the cavity, $a$ and $a^\dagger$ are the
annihilation and creation operators for the quantum oscillators,
and $\omega_c$ is its frequency. Therefore, we have
$$\frac{2\pi}{\phi_0}\int_{l_m} {\bf A}_m\cdot d {\bf l}_m
=\frac{2\pi }{\phi_0} \int_{l_m} {\bf A}^\phi_m\cdot d {\bf l}_m +
g(a+a^{\dagger}),$$ where $\phi_0=\pi\hbar/e$ being the flux
quantum, $g=2e \hat{\epsilon} \cdot {\bf l}/\sqrt{2\varepsilon
\omega_c V\hbar}$ is the coupling constant between the junctions
and the cavity, with $l$ the thickness of the insulating layer in
the junction. The closed path integral of the ${\bf A}^\phi$ gives
rise to the magnetic flux: $\oint_C {\bf A}^\phi d {\bf l} =\Phi$.
Similar to that in Ref. \cite{zhucavity}, setting
$\pi\Phi/\phi_0=\omega t$ with $\omega_c-\omega=\delta\ll\omega$,
the Hamiltonian (\ref{h1}) reads
\begin{eqnarray}  \label{hcoupling}
H=E_{c} (n-\bar{n})^2 \sigma_z
-\frac{E_J}{2}\left(\sigma^{\dag}e^{-i[\frac{g}{2}
(a+a^\dagger)+\omega t]} +\text{H.c.}\right).
\end{eqnarray}

Consider that $N$ such qubits are located within a single-mode
cavity. To a good approximation, the whole system can be considered
as $N$ two-level systems coupled to a quantum harmonic oscillator
\cite{Almaas}. Setting the qubits at their degeneracy points, the
system considered here can then be described by the Hamiltonian $H=
H_0+H_{int},$ where
\begin{subequations}
\begin{eqnarray}
\label{H_0} H_0=\hbar\omega_c \left(a^\dagger a+\frac{1}{2}\right),
\end{eqnarray}
\begin{eqnarray}
\label{Hint} H_{int}=-\frac{E_J}{2}\sum_{j=1}^{N}\left(
\sigma_j^{+}e^{-i[\frac{g}{2} (a+a^\dagger)+\omega
t]}+\text{H.c.}\right).
\end{eqnarray}
\end{subequations}
The spin notation is used for the qubit $j$ with Pauli matrices
$\{\sigma^{x}_j,\sigma^{y}_j,\sigma^{z}_j \}$, and
$\sigma_j^{\pm}=(\sigma^x_j \pm i\sigma^y_j)/2$. For simplicity,
we have also assumed the same $E_{c}$ and $E_{J}$  for all
qubits. Expanding the Hamiltonian (\ref{Hint}) to the first order
of $g$ in the Lamb-Dicke limit and under the rotating wave
approximation as well as in the interaction picture
$U_0=\exp(-iH_0 t)$, the Hamiltonian is reduced to
\begin{eqnarray}
\label{hint} H_{int}=\frac{igE_J}{4}\left(a^{\dag}e^{i\delta
t}-ae^{-i\delta t}\right)J_x,
\end{eqnarray}
where $J_x=\sum_{j=1}^n\sigma^x_j$. The time-evolution operator for
Hamiltonian (\ref{hint}) can be expressed as
\begin{eqnarray}
\label{u}
U(t)&=&\exp\left\{\left[\int_0^{t}B^{*}_{(t)}dB_{(t)}\right]J _x^2
\right\}\nonumber\\
&\quad&\times\exp\left[iB^{*}_{(t)}aJ_x\right]\exp\left[iB_{(t)}a^\dag
J_x\right],
\end{eqnarray}
where $B_{(t)}=(gE_J/4\hbar\delta)(1-e^{i\delta t})$. Setting
$\delta T=2k\pi$ ($k=1,2,\cdot\cdot\cdot$) leads to
\begin{eqnarray}\label{jx}
U(\gamma)=\exp(i\gamma J _x^2)
\end{eqnarray}
with
\begin{eqnarray}\label{T}
\gamma=\left(\frac{gE_J}{4\hbar\delta}\right)^2\delta T.
\end{eqnarray}
Interestingly, this $U(\gamma)$-operation is insensitive to the
thermal state of the cavity mode as  the related influence
represented by the last two exponents in Eq. (\ref{u}) is
completely removed.

It is also notable that, when only two qubits are considered in Eq.
(\ref{jx}), it is straightforward to check that $U(\gamma)$ is a
nontrivial two-qubit unconventional geometric phase gate
~\cite{zhuunconventional},  where the phase $\gamma$  satisfies the
relation $\gamma=\gamma_g+\gamma_d=-\gamma_g$ (i.e.,
$\gamma_d=-2\gamma_g$), with $\gamma_g$ and $\gamma_d$ being
respectively the geometric  and dynamic phases accumulated in the
evolution; this unconventional geometric phase shift  still depends
only on global geometric features and is robust against random
operation errors \cite{zhuunconventional}, thus the high-fidelity of
the two-qubit operation may be experimentally achieved. For example,
as an entangling operation gate,  it can entangle two qubits from a
separable state to a fully entangled EPR-state
\begin{eqnarray}
|00\rangle\stackrel{{U(\gamma)}}\longrightarrow\frac{1}{\sqrt{2}}(|00\rangle+i|11\rangle),
\end{eqnarray}
once we set $\gamma=\pi/8$. More generally, the operation (\ref{jx})
can be used to generate multipartite entangled GHZ state with an
extended unconventional geometric phase shift scenario in this
system \cite{zhucavity}. In addition, this solid-state architecture
may also provide an alternative geometric approach to construct
quantum error correcting code \cite{zhucavity}.

In most current efforts, the universal quantum computation is
achieved with sequences of controlled interactions between selected
qubits. Being significantly different from these efforts,
Raussendorf and Briegel \cite{oneway} proposed a new kind of
scalable quantum computation, namely the one-way quantum
computation, which constructs quantum logic gates by single-qubit
measurements on cluster states. The distinct advantage of one-way
computing strategy lies in that it separates the processes of
generating entanglement and executing computation. So one can
tolerate failures during the generation process simply by repeating
the process, provided that the failures are heralded. Due to its
novel application in quantum computing, the generation of cluster
states has also been proposed in the context of the cavity QED with
atomic qubits \cite{gqed} and superconducting qubits \cite{gsquid}.
However, all these generation schemes are of the "step by step"
nature, which means that the time needed for generating the cluster
states is determined by the number of the qubits, thus the
scalability of these schemes is unlikely good. Very recently,
Tanamoto \emph{et al}. \cite{tanamoto} proposed a new scheme to
generate the cluster state of superconducting qubits by only one
step. However, since the inter-qubit interaction was capacitively
coupled in the scheme \cite{tanamoto}, where each qubit works far
away from the degeneracy point, the decoherence time of a single
qubit is much shorter than that at the degeneracy point. In
addition, the capacitive inter-qubit coupling is fixed
\cite{chargeent}, thus it is difficult to prepare the initial state
for each qubit. You \emph{et al}. \cite{youcluster} proposed another
scenario to improve the performance of generating cluster states by
introducing an inductive inter-qubit coupling. We note that both
schemes need significant resources to couple different qubits
\cite{tanamoto,youcluster}.

As a direct and useful application of the multi-qubit operator
(\ref{jx}), we here present an efficient way for generating the
multipartite cluster states. In the present scalable solid-state
system, the effective long range couplings among qubits are mediated
by the cavity field, and thus no auxiliary devices are needed. The
operator (\ref{jx}) is equivalent to
\begin{eqnarray}\label{u2}
U(\gamma)_2=\exp\left(i2\gamma\sum_{j>i=1}^n\sigma_x^i\sigma_x^j\right),
\end{eqnarray}
up to an overall phase factor. Since the two operators (\ref{u1})
and (\ref{u2}) commute with each other, we can perform the two
corresponding operations with the time intervals $t_1$ and $T$
sequentially to obtain the wanted unitary operation \cite{note}.
Setting
\begin{eqnarray}\label{c1}
2E(\Phi)t_1=(N-1)\hbar\gamma,
\end{eqnarray}
we have the total evolution operator as
\begin{eqnarray}\label{uc}
U(t_1+T)=\exp\left[i8\gamma\left(\sum_{j>i=1}^N\frac{1+\sigma_x^i}{2}\frac{1+\sigma_x^j}{2}\right)\right].
\end{eqnarray}

The initial state of each charge qubit can be prepared as
$$|0\rangle_i=\frac{1}{\sqrt{2}}\left(|-\rangle_i + |+\rangle_i\right),$$
where $|\pm\rangle_i=(|0\rangle_i \mp |1\rangle_i)/\sqrt{2}$ are
eigenstates of $H_i=-E(\Phi)\sigma_x^{(i)}$ with eigenvalues $\pm
E(\Phi)$. When the condition $\gamma=(2n+1)\pi/8$
 is satisfied, the generated cluster state is
\begin{equation}
\label{cluster} \frac{1}{2^{N/2}}\bigotimes_{i=1}^N\left(|-\rangle_i
(-1)^{N-i}\prod_{j=i+1}^N\sigma_x^{(j)} + |+\rangle_i\right),
\end{equation}
which is a highly entangled state. The operator $\sigma_x^{(j)}$
acts on the states $|\pm\rangle$ of the qubits $j=i+1,\dots,N$, with
$i=1,2,\dots,N-1$; this is due to the inter-qubit coupling mediated
by cavity field is of the long-range nature. The above condition can
be satisfied whenever
\begin{equation}
\label{deltag} \delta=\frac{gE_J}{\hbar}\sqrt{\frac{k}{2n+1}}.
\end{equation}
Correspondingly, the requirement (\ref{c1}) may be expressed as
\begin{equation}\label{single}
t_1=\frac{(N-1)(2n+1)\pi\hbar}{16E(\Phi)}.
\end{equation}
In the present proposal for producing cluster states, an effective
anisotropic direction is along the $x$-axis, rather than the
$z$-direction for the standard Ising model \cite{oneway}. Now the
cluster state is represented using the eigenstates of $\sigma_x$,
and correspondingly, the single-qubit projective measurements are
performed on the eigenstates of $\sigma_z$.

Comparing with the existing schemes for generating cluster states,
our scheme possesses likely the following  advantages.

(1) As a macroscopic analogy of atomic quantum computing and
control, the present one provides strong inhibition of spontaneous
emission and suppresses the decoherence caused by the external
environment.  The positions of qubits in the cavity are fixed and
thus they can be easily and selectively addressed. It seems easy  to
scale up to large number of qubits with the present one, and the
control and measurement techniques are more advanced for this
system.

(2) The present scheme works at the degeneracy point, where the
qubit has a longer decoherence time. The initialization and
operation (\ref{u1}) for the qubits can be easily achieved. After
generating the cluster state, no external magnetic flux is applied
and the inter-qubit coupling is also switched off. This is
convenient for implementing one-way quantum computation via local
single-qubit measurements, which can be more efficiently
implemented, e.g., using a single-electron transistor coupled to the
charge qubit \cite{Makhlin}.

(3) Our scheme is quite simple and feasible in comparison with that
addressed in  Ref. \cite{zhucavity}, where the superconducting qubit
was formed by two symmetric superconducting quantum interference
devices connected by a $\pi$ junction.  Each qubit in Ref.
\cite{zhucavity} is controlled via two different frequencies of
microwave, while we here only need one of them.

(4) It is not a "step by step" one, thus its scalability is better
than those in Refs. \cite{gqed,gsquid}. In our scheme,  since the
time  needed for a single qubit operation is negligibly short in
comparison with the time needed for the multipartite collective
operation, which is independent of the number of qubits involved,
the total time needed for generating the cluster states is almost
independent on the number of the qubits.

(5) In our scheme, the couplings among different qubits are mediated
by the cavity field, and thus no auxiliary devices are needed,
noting that auxiliary devices would increase the complexity of the
circuits and inevitably introduce additional uncontrollable noises.
Besides, the coupling of any qubit to the cavity could be easily
switched on and off via the control of the external gate voltage and
the flux of the microwave. In addition, the cavity-qubit coupling is
insensitive to the thermal state of the cavity mode by removing the
influence of the cavity mode via the periodical evolution.

Before concluding the paper, we briefly address the experimental
feasibility of the proposed scheme with the parameters already
available in current experimental setups. Suppose that the quality
factor of the superconducting cavity is $Q=1\times10^{6}$
\cite{Day}, for the cavity with $\hbar\omega_c=30\mu ev$
 ($\omega_c=30$GHZ)\cite{Wallraff}, the cavity decay time is
$\tau=Q/\omega_c\approx33$$\mu$s. The decoherence time of qubit
without the protection of the cavity is $T_d\approx0.5\mu$s
\cite{Vion}.From Eq. (\ref{deltag}), we estimate
$\delta\approx0.6$GHZ for $g=10^{-2}$ \cite{Almaas}, $E_J=40\mu$ev
\cite{chargeent,cg}, $k=1$ and $n=0$. The time for single  qubit
 rotation without cavity is $t_1\approx3.3\times(N-1)$ps for
$\gamma=\pi$/8 from Eq. (\ref{single}),  and  $T\approx10$ns from
Eq. (\ref{T}). Thus the total manipulation time
$t_k=t_1+T\approx10$ns, which is much shorter than the cavity decay
time $\tau$ and the decoherence time of qubit $T_d$. With the vacuum
Rabi frequency $\Omega=15$MHz and the lifetime of the qubit
$\gamma=2\mu$s, the strong coupling limit can be readily fulfilled
($\Omega^2\tau\gamma \sim 10^4\gg1$). So, with the properly chosen
parameters, both Lamb-Dicke and strong coupling limits can be
fulfilled simultaneously.

In summary, we have proposed a new simple scheme for implementing
the multi-qubit operations with superconducting charge qubits
coupled to a microwave cavity mode. The quantum operations depend
only on global geometric features and are insensitive to the thermal
state of the cavity mode, and thus it may result in high-fidelity
quantum information processing. In particular, we have illustrated
how to generate the highly entangled cluster state more efficiently
in the present solid-state system without auxiliary devices, which
is promising for realizing one-way quantum computation.

We thank S.-L. Zhu and L.-X. Cen for many helpful discussions.
This work was supported by the RGC of Hong Kong under Grant No.
HKU7045/05P, the URC fund of HKU, the NSFC under Grant No.
10429401, and the State Key Program for Basic Research of China
(2006CB0L1001).


\begin{thebibliography}{99}
\bibitem{note} Note that in the absence of the external AC magnetic flux,
the system is in the large-detuning limit (the frequency detune
between the cavity mode and the qubit is much larger than their
coupling constant) and the interaction between the cavity mode and
the qubit can be neglected \cite{liu}. In other words, by the
sequential application of the static and AC flux, we can effectively
decouple and then couple the interaction, and thus the operation
$U(t)_1$ plus $U(\gamma)_2$ is implemented. It is also noted that
another interesting couple/decouple mechanism is presented in Ref.
\cite{wei}, where a frequency-matching technique is adopted.
\end{thebibliography}
\end{document}